\documentclass[a4paper,12pt]{article}
\pdfoutput=1 

\usepackage{jheppub} 

\usepackage{fancyhdr}
\usepackage{graphicx}

\usepackage{braket}
\usepackage{slashed}

\newcommand{\be}{\begin{equation}}
\newcommand{\ee}{\end{equation}}
\newcommand{\bea}{\begin{eqnarray}}

\newcommand{\eea}{\end{eqnarray}}

\newcommand{\Rmnum}[1]{\expandafter\@slowromancap\romannumeral #1@}

 \title{\boldmath Isospin-violating dark matter from a double portal}

\author[a]{Genevi\`eve B\'elanger,}
\author[a]{Andreas Goudelis,}
\author[b]{Jong-Chul Park,}
\author[c]{Alexander Pukhov}

\affiliation[a]{LAPTh, Universit\'e de Savoie, CNRS, 9 Chemin de Bellevue, B.P.\
110,
F-74941 Annecy-le-Vieux, France
}
\affiliation[b]{Department of Physics, Sungkyunkwan University, Suwon 440-746, Korea}
\affiliation[c]{Skobeltsyn Institute of Nuclear Physics, Moscow State University, Moscow 119992, Russia}
\emailAdd{belanger@lapth.cnrs.fr}
\emailAdd{andreas.goudelis@lapth.cnrs.fr}
\emailAdd{log1079@gmail.com}
\emailAdd{pukhov@lapth.cnrs.fr}

\abstract{We study a simple model that can give rise to isospin-violating interactions of Dirac fermion asymmetric dark matter to protons and neutrons through the interference of a scalar and U(1)$'$ gauge boson contribution. The model can yield a large suppression of the elastic scattering cross section off Xenon relative to Silicon thus reconciling CDMS-Si and LUX results while being compatible with LHC findings on the 126 GeV Higgs, electroweak precision tests and flavour constraints.  }

\begin{document}
\maketitle
\flushbottom

\section{Introduction} \label{sec:intro}
Several dark matter (DM) direct detection experiments have observed an excess of events which, when interpreted as dark matter signals, would imply dark matter masses below the electroweak scale.
Such experiments include DAMA/LIBRA~\cite{Bernabei:2010mq}, CoGeNT~\cite{Aalseth:2010vx,Aalseth:2012if}, CRESST II~\cite{Angloher:2011uu}, and more recently CDMS II Si~\cite{Agnese:2013rvf}.
DAMA/LIBRA has observed an annual modulation signal, CRESST and CDMS II Si have reported unmodulated ones, while CoGeNT has published results on both.
The best-fit to the three events observed by CDMS-Si are given by a WIMP of mass $8.6$ GeV and elastic scattering cross-section of $2\times 10^{-5} \rm{pb}$, a range also preferred by CoGeNT. Similarly,  the CRESST-II results are compatible with a WIMP of a mass 10-40 GeV and a cross section in the range $10^{-6}-10^{-4}{\rm pb}$ while DAMA/LIBRA favours a larger cross section (few $10^{-4}{\rm pb}$).
On the other hand, other experiments, notably  XENON10~\cite{Angle:2011th},  XENON100~\cite{Aprile:2012nq}, and recently LUX~\cite{Akerib:2013tjd} have derived exclusion limits   that are incompatible with  these signals for most  of the preferred area in the mass/cross section plane when all results are interpreted in terms of spin independent (SI) interactions of equal strength on protons and neutrons.

Spin independent interactions  that are isospin violating and specifically with a ratio of the amplitude for neutrons and protons
$f_n/f_p \simeq -0.7$ \cite{Chang:2010yk, Feng:2011vu, Frandsen:2013cna} have been suggested as a way to reconcile positive results obtained with light nuclei and  exclusion limits obtained with Xenon. Indeed for this specific ratio of amplitudes, the scattering cross-section off  Xenon is strongly suppressed
due to the destructive interference between the amplitudes on neutrons and protons, while that for lighter nuclei like Si is suppressed much more mildly. General suppression factors for isospin violating interactions relative to the isospin conserving case for various elements can be found in Ref.~\cite{Kopp:2011yr, Feng:2013vaa}. Such isospin violating interactions would therefore allow the reconciliation of the CDMS-Si (and to a certain extent the CoGeNT) result with the exclusion bounds coming from Xenon detectors.  Note however that the corresponding tension with the DAMA result, obtained with NaI, cannot be fully resolved.

Constructing a realistic particle physics model that can reproduce the amplitudes with the required ratio and leading to a sufficiently large scattering cross-section while satisfying other dark matter and collider constraints is a challenge
(for some attempts, see e.g. \cite{Cline:2011zr,Okada:2013cba,Kang:2010mh,Gao:2011ka}).
First, we observe that the Higgs exchange leads to nearly equal amplitude for protons and neutrons, therefore the Higgs cannot be the sole mediator of interactions with nuclei.
Second, if the dark matter interacts with  the Higgs it would lead to invisible decays of the latter unless its coupling to the  Higgs is  suppressed. The discovery of a Higgs boson with a mass of 126 GeV at the LHC~\cite{Aad:2012tfa,Chatrchyan:2012ufa} and the measurements of its properties  constrain the
invisible decay width to be  below 30\%~\cite{Belanger:2013kya, Belanger:2013xza, Giardino:2013bma} and thus limit the strength of the interactions with nucleons~\cite{Djouadi:2011aa,He:2011gc,Belanger:2013kya,Greljo:2013wja}. The spin independent interactions with nuclei must therefore receive important contributions from other particles, for example an extra scalar or an extra gauge boson (the latter contributing only if dark matter is not self-conjugate). The first possibility was investigated in~\cite{He:2013suk}   and the second in~\cite{Frandsen:2011cg}  in models with scalar dark matter.
In this work we consider another option, that of a
Dirac fermion dark matter candidate which  can interact with a light new gauge boson (a $Z'$ or `dark photon') with couplings $f_p\gg f_n$.
In order to achieve the needed amount of isospin violation to  suppress the spin independent interaction with Xenon while not affecting too drastically the interaction with Si, we make use of the interference between the Higgs and vector boson exchanges.
Since only one of the two (dark matter or anti-dark matter) components possesses the correct-sign coupling to the new gauge boson that can lead to a destructive interference between Higgs and vector boson exchange contributions, such an interference requires some dark matter asymmetry.

In what follows, the general picture that will emerge from the requirements on elastic scattering cross sections is that the relic density must be driven  by the asymmetric component and thus a value compatible with PLANCK results  can be easily obtained  by appropriately adjusting the  initial asymmetry. This also implies that the relic density component resulting from thermal freeze-out must be very small. To achieve this, dark matter annihilation can be enhanced by the quasi-resonant $s$-channel exchange of  a $Z'$ boson.
Although we do not attempt to explain the origin of the asymmetry, such setups are interesting since they could be related to the same mechanism that leads to a
small excess of matter over anti-matter in the early universe. The excess of DM over anti-DM  in the early universe will be taken of the same order as the baryonic asymmetry thus naturally leading to a relic density of DM
of the same order (a factor of 5 higher) than that of ordinary matter, for a recent review see Ref.~\cite{Petraki:2013wwa}.
In this model,  limits on invisible $Z$ and Higgs decays,   constraints from Higgs searches at colliders,  from Kaon and B physics, and from electroweak precision measurements can all be  satisfied.

The outline of the paper goes as follows: In section \ref{sec:modelpspace} we present the model and some key relations. In section \ref{sec:psconstraints}, we discuss the parameter space of the model and the constraints it is subject to. Then, in section \ref{sec:supressionmec} we analytically explain the mechanism that allows us to reconcile the direct detection results of CDMS-Si with those of Xenon detectors and illustrate it with concrete numerical examples. In section \ref{sec:scans} we perform a comprehensive scan over the model's parameter space and locate regions where the CDMS-Si result can be reproduced without contradicting the null results from XENON100 and LUX. Finally, we conclude in section \ref{sec:conclusions}. In appendix \ref{app:interactions} we provide for convenience the most important couplings of our model.

\section{Model and parameter space}\label{sec:modelpspace}
In this section we briefly present the various ingredients of our model,
provide some key relations that will be of importance in the following, and describe the model's parameter space.
\subsection{The model}\label{themodel}

The model we consider in this work consists of the Standard Model (SM) extended by an additional U(1)$_X$ gauge group factor, a hidden sector
containing a Dirac fermion $\psi$ that is
neutral under ${\rm SU(3)}_c\times {\rm SU(2)}_L \times {\rm U(1)}_Y$ but charged under U(1)$_X$ and will subsequently play the role of a dark matter candidate,
as well as a real singlet scalar field $S$. The hidden sector can couple to the SM sector
through a ``double portal'' interaction: a mixing of the usual Higgs doublet and the $S$ singlet in the scalar potential, a ``Higgs portal''
interaction~\cite{Binoth:1996au, Schabinger:2005ei}, and a kinetic mixing between U(1)$_X$ and U(1)$_Y$~\cite{Okun:1982xi, Holdom:1985ag, Dienes:1996zr, Huh:2007zw, Chun:2008by, Chun:2010ve,Mambrini:2010dq,Belanger:2011ww, Park:2012xq}. The Lagrangian we adopt, including both mixings, reads
\begin{align} \label{Lagrangian}
{\cal L} = {\cal L}_{SM} & - {1\over2} \sin\epsilon\, \hat{B}_{\mu\nu} \hat{X}^{\mu\nu} -\frac{1}{4}\hat{X}_{\mu\nu}\hat{X}^{\mu\nu}  + {1\over2} m_{\hat{X}}^2 \hat{X}^2 + y_\psi S \bar{\psi}\psi + g_{X} \hat{X}_\mu \bar{\psi}\gamma^\mu \psi \nonumber\\
&- \lambda_{SH} S^\dagger S H^\dagger H + \frac{1}{2}\mu_S^2 S^\dagger S - \frac{1}{4} \lambda_S (S^\dagger S)^2 + \frac{1}{2}\mu_H^2 H^\dagger H - \frac{1}{4} \lambda_H (H^\dagger H)^2   \,,
\end{align}
where the hidden gauge boson mass $m_{\hat{X}}$ can result from the spontaneous breaking of U(1)$_X$ or through some  alternative to the Higgs
mechanism, such as the Stueckelberg mechanism~\cite{Stueckelberg:1938zz, Kors:2004dx}.
In the SM sector, the mass of the $\hat{Z}$ gauge boson is $m_{\hat{Z}}$ and the gauge couplings are denoted by $\hat{g}=\hat{e}/s_{\hat{W}}$
and $\hat{g}'= \hat{e}/c_{\hat{W}}$.

The Lagrangian \eqref{Lagrangian} contains both kinetic and mass off-diagonal terms mixing the $\hat{B},\hat{W}_3$ and $\hat{X}$ gauge bosons.
The passage to the physical $(A, Z, Z_X)$ basis can be performed by diagonalizing away the kinetic and mass mixing terms through the following transformation:
\begin{align} \label{transformation}
\hat{B} & = c_{\hat{W}} A - (t_\epsilon s_\xi+ s_{\hat{W}} c_\xi) Z + (s_{\hat{W}} s_\xi-t_\epsilon c_\xi) Z_X \,, \nonumber\\
\hat{W}_3 & = s_{\hat{W}} A + c_{\hat{W}} c_\xi Z - c_{\hat{W}} s_\xi Z_X \,, \nonumber\\
\hat{X} & = {s_\xi \over c_\epsilon} Z + {c_\xi\over c_\epsilon} Z_X \,,
\end{align}
where the rotation angle $\xi$ is determined by
\begin{equation} \label{t2csi}
\tan2\xi = - {m_{\hat{Z}}^2 s_{\hat{W}} \sin2\epsilon \over
               m_{\hat{X}}^2 - m_{\hat{Z}}^2
               (c^2_\epsilon-s^2_\epsilon s_{\hat{W}}^2) }
\end{equation}
and the weak mixing angle $s_{\hat{W}}$ is very close to the physical value $s_W$ due to the stringent constraint on the parameter $\rho \equiv m_W^2/m_Z^2 c_W^2$, $\rho = 1.0004^{+0.0003}_{-0.0004}$ \cite{Beringer:1900zz}.
Then, the masses of the $Z$ and $Z_X$ gauge bosons are redefined as,\footnote{One can find a detailed analysis on the kinetic mixing part in Ref.~\cite{Chun:2010ve}.}
\begin{align}
m_Z^2 &= m_{\hat{Z}}^2(1+s_{\hat{W}} t_\xi t_\epsilon) \,,  \label{mz} \\
m_X^2 &=
   { m_{\hat{X}}^2 \over c_\epsilon^2 (1+s_{\hat{W}} t_\xi t_\epsilon)} \,.
\label{mx}
\end{align}
On the other hand, the mass of the physical $W$ boson remains unaffected by the transformation~\eqref{transformation},
\begin{equation}
m_W^2 = m_{\hat{W}}^2 = m_{\hat{Z}}^2 c_{\hat{W}}^2\,,
\end{equation}
which means that the $\rho$ parameter can be written as
\begin{equation}
\rho  = \frac{c_{\hat{W}}^2}{(1+s_{\hat{W}}^2 t_\xi t_\epsilon) c_W^2}.
\end{equation}
As pointed out in Ref.~\cite{Babu:1997st}, the photon coupling also remains unchanged. This fact can be used to deduce the relation
\begin{equation}
c_W^2 s_W^2 = \frac{c_{\hat{W}}^2 s_{\hat{W}}^2}{1 + s_{\hat{W}} t_\xi t_\epsilon }
\end{equation}
which leads to
\begin{equation}
\rho = \frac{s_W^2}{s_{\hat{W}}^2}.
\label{eq:rhopar}
\end{equation}

Passing to the scalar sector of the model now, upon electroweak symmetry breaking we can as usual expand the scalar doublet and singlet that, in the unitary
gauge, take the form
\begin{align}\label{Higgs}
H = \frac{1}{\sqrt{2}}
\left( \begin{array}{cc}
0 \\ v+h
\end{array} \right)\,,\quad
S = \frac{1}{\sqrt{2}} (v_S + s)\,,
\end{align}
where $v=246$ GeV. Then the mass of the hidden fermion $\psi$ is $m_\psi = y_\psi v_S/\sqrt{2}$. The squared mass matrix of the Higgs sector is in turn given by
\begin{align}\label{MassMatrix}
\mathcal{M}_{sh}^2 =
\left( \begin{array}{cc}
\lambda_S v_S^2/2 & \lambda_{SH} v v_S \\ \lambda_{SH} v v_S & \lambda_H v^2/2
\end{array} \right)\,,
\end{align}
where we have used the minimization conditions of the Higgs potential:
\begin{align}\label{minimization}
\mu_S^2 - \frac{1}{2}\lambda_S v_S^2 - \lambda_{SH} v^2 = 0\,,\quad
\mu_H^2 - \frac{1}{2}\lambda_H v^2 - \lambda_{SH} v_S^2 = 0\,
\end{align}
to eliminate the parameters $\mu_H^2$ and $\mu_S^2$. The eigenvalues of the mass matrix \eqref{MassMatrix}, corresponding to the physical scalar
states $h_1$ and $h_2$, are
\begin{align}\label{HiggsMassses}
m_{h_1, h_2}^2 = \frac{1}{4}\lambda_H v^2 + \frac{1}{4}\lambda_S v_S^2
\mp \sqrt{\left(\frac{1}{4}\lambda_H v^2 - \frac{1}{4}\lambda_S v_S^2 \right)^2
+(\lambda_{SH} v v_S)^2}
\end{align}
with
\begin{align}\label{rotation}
\left( \begin{array}{cc} h_1\\ h_2 \end{array}\right) =
\left( \begin{array}{cc}
c_\alpha & -s_\alpha \\ s_\alpha & c_\alpha
\end{array} \right)
\left( \begin{array}{cc} s\\ h \end{array}\right)\,,
\end{align}
where the rotation angle $\alpha$ is given by
\begin{align}\label{t2alpha}
\tan {2\alpha} = \frac{4 \lambda_{SH} v v_S}{\lambda_H v^2 - \lambda_S v_S^2}\,.
\end{align}
The couplings of scalar particles to fermions are modified as
\begin{align}
& g_f^{h1} = -s_\alpha y_f/\sqrt{2}\,\,,\quad
g_\psi^{h1} = c_\alpha y_\psi/\sqrt{2}\,, \label{h1Couplings} \\
& g_f^{h2} = c_\alpha y_f/\sqrt{2}\,,\quad\quad
g_\psi^{h2} = s_\alpha y_\psi/\sqrt{2}\,,  \label{h2Couplings}
\end{align}
where $y_f = \sqrt{2} m_f/v$ and $y_\psi = \sqrt{2} m_\psi/v_S$.

In Appendix \ref{app:interactions}, we list for convenience the full set of $W$, $Z$ and $Z_X$ gauge boson couplings resulting from the Lagrangian \eqref{Lagrangian}
that are of relevance for our analysis in the physical field basis, as well as the expressions for the triple scalar couplings in the physical Higgs boson basis.

\subsection{Parameter Space}\label{sec:pspace}

The model, as defined from Eq.\eqref{Lagrangian}, can be described by a set of $11$ parameters
\begin{equation}\label{LAGparameters}
\ m_{\hat{Z}}, \ m_{\hat{W}}, \ m_{\hat{X}}, \ \sin\epsilon, \ g_{X}, \   y_\psi, \ \lambda_{SH}, \ \lambda_S, \ v_S, \ \lambda_H, \ v\,,
\end{equation}
the last two of which are already present in the SM. In practice, we can use the relations presented in the previous section to exchange some
of these parameters with more physically meaningful ones. Hence, in what follows we will rather be working in the space defined by the following set of parameters
\begin{equation}\label{PHYSparameters}
\ m_Z, \ m_W, \ \ m_{Z_X}, \ \epsilon, \ \ g_X, \ y_\psi, \ m_\psi, \ \rho,  \ m_{h_1}, \ m_{h_2}, \ \alpha\,,
\end{equation}
where $m_{Z, W, Z_X}$ are the masses of the physical $Z$, $W$ and $Z_X$ bosons respectively and $m_{h_{1, 2}}$ are the masses of the physical Higgs bosons for which according to the notations in Eq.~\eqref{HiggsMassses} we have $m_{h_2} > m_{h_1}$.
Note that by using $\rho$ as a free parameter of the model, and by letting it vary within its experimental bounds, we automatically ensure that
all the results we will present in the following are compliant to the $\rho$ parameter constraint.

We should point out that in this work, we will not examine the full range of allowed values for the parameter space.
Motivated by the CDMS-Si excess which is compatible with low-mass dark matter,  we will focus in particular on low values
for the dark matter candidate mass $m_\psi$. The rest of the parameters will in turn be chosen so as to satisfy the experimental constraints,
to be described in the following section, as well as to reproduce the direct detection effects we are interested in. We should also however
stress that part of the discussion that follows has a scope extending well beyond any attempt to reconcile the CDMS-Si and
LUX results. We will further clarify this point later on.

\section{Constraints}\label{sec:psconstraints}
Our setup is subject to a series of constraints coming from different sources, which interestingly affect in a distinct manner the various sectors
of the model: low-energy observables, collider bounds as well as cosmological measurements. In this section we describe these constraints and the
way they are accounted for in our analysis.
\subsection{Constraints on the gauge sector}\label{sec:gaugesector}

A first set of observables stemming from low-energy and LEP measurements allow us to constrain the gauge sector of the model and its interactions
to fermions. First, electroweak precision tests (EWPT) allow us to set limits on combinations of $(m_{Z_X}, \epsilon)$ values. Comprehensive analyses
of such constraints have been performed in \citep{Kumar:2006gm,Chang:2006fp}. Here we adopt the approximate limit
\begin{equation}
\left( \frac{\tan\epsilon}{0.1} \right)^2 \left(\frac{250 \ \mathrm{GeV}}{m_{Z_X}}\right)^2 \leq 1\,.
\end{equation}
Second, the $\rho$ parameter also imposes a constraint on the gauge sector, which in our choice of parameter space basis can be satisfied by simply choosing $\rho \in (0.9992,1.0016)$, i.e. a $3\sigma$ interval around the central value.

The mixing among the two U(1)'s moreover modifies the physical $Z$ boson decay modes. In particular, when the $\psi$ DM candidate is light enough, as is the case
in this work, the $Z$ can then decay into pairs of DM particles.
The most stringent constraints on the $Z$ total width come from precision measurements on the
$Z$ pole performed at LEP \cite{Z-Pole} that sets the uncertainty in the total $Z$ width at $1.5$ MeV (at $68\%$CL),
which also fixes the maximally allowed decay width into exotic modes. We impose the condition
\begin{equation}
\Gamma(Z \rightarrow \psi\bar{\psi}) < 3 \times 0.0015 \ \ \mathrm{GeV}
\label{eq:Zwidth}
\end{equation}
i.e. we again demand for our results to be compatible with the experimental measurements within $3\sigma$.

Other constraints on a new light gauge boson arise from low energy neutral currents, atomic parity violation, the muon anomalous magnetic moment~\cite{Chun:2010ve} or from flavour constraints~\cite{Williams:2011qb,Frandsen:2011cg}. However, in this model where the coupling of $Z_X$ to standard model fermions is only introduced through mixing with the $Z$, these constraints are easily avoided after  taking into consideration the EWPT and LEP constraints discussed above.

\subsection{Constraints on the scalar sector}\label{sec:scalarsector}

A crucial and less studied constraint arises in the scalar sector of the model after the LHC discovery of a
Higgs-like particle. As a first remark, let us note that with the particle content considered in this paper, the production modes of the Higgs boson are
essentially identical to the Standard Model ones (given the strong constraints on the gauge boson sector we expect that Vector Boson Fusion should
not be significantly modified). In a series of recent studies \cite{Belanger:2013kya,Belanger:2013xza} it has been shown that under these
circumstances, the total branching ratio of the Higgs boson into invisible decay modes has to obey
\begin{equation}\label{eq:higgsinvcons}
BR(h \rightarrow inv) \lesssim 0.3 \ .
\end{equation}
We should note that by ``invisible'' here we do not only mean decays into \textit{actually} invisible
(i.e. $E_T^{Miss}$-only) final states. Instead, under the general label of ``invisible'' decays we should include all possible decay modes of the Higgs boson that are
not accounted for in experimental studies. Denoting the SM-like Higgs boson by $h$, in our setup we have three such possible
modes depending on the mass hierarchy of the involved particles: $h \rightarrow Z_X Z_X$, $h \rightarrow \psi\psi$ and $h \rightarrow h_1 h_1$  when $h$ coincides with $h_2$. In
anticipation of the analysis that will follow, we point out that for the parameter ranges that we will study, the first of these
decay modes turns out to be negligible. The other two modes, however, can be particularly important and will crucially affect the mass range of the non-SM like
Higgs boson. In the subsequent analysis, we will demand that the total $BR(h \rightarrow inv)$ does not exceed $30\%$.

\begin{figure}[t!]
\begin{center}
\includegraphics[width=0.70\textwidth]{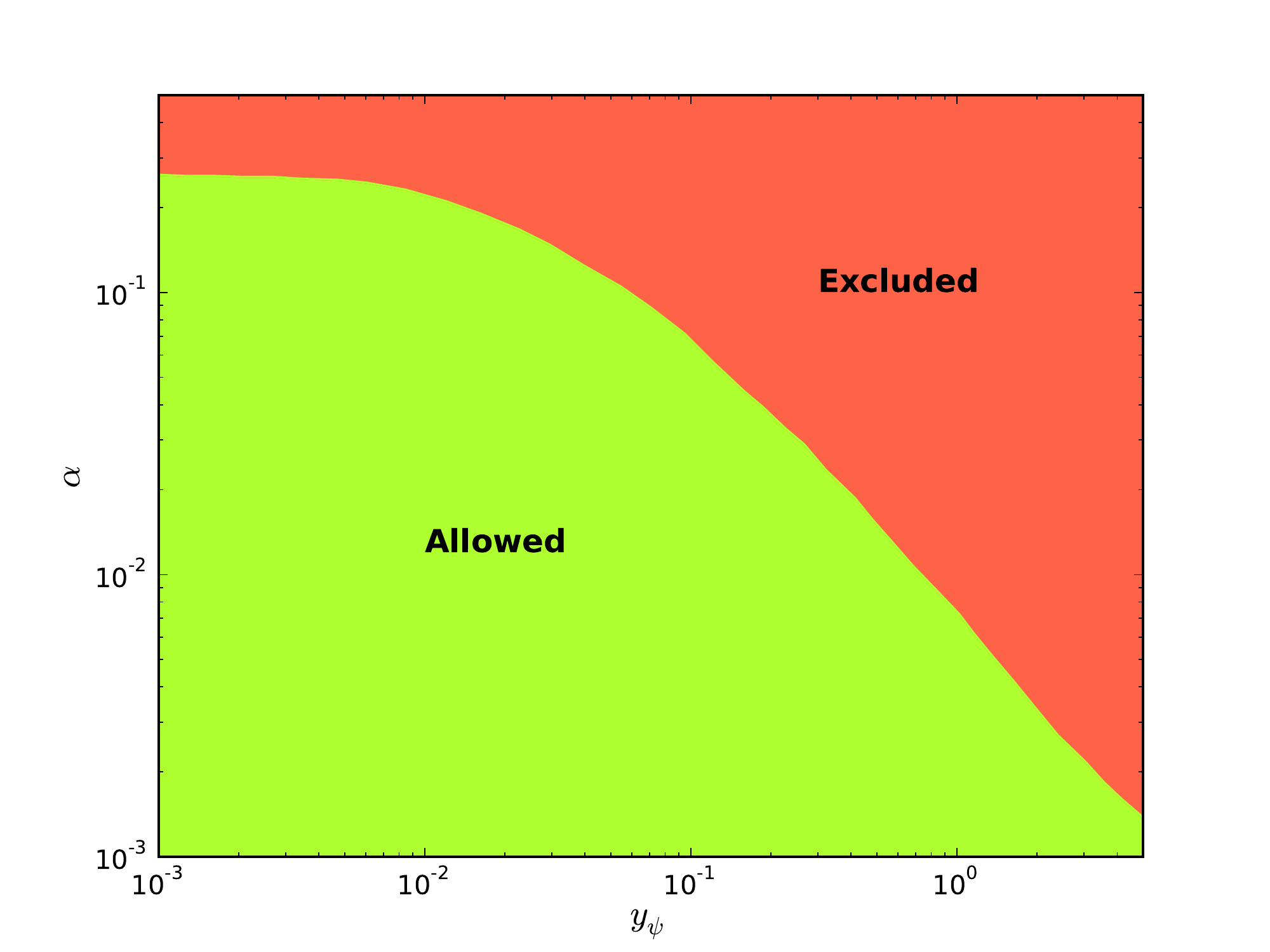}
\caption{Exclusion bounds from non-standard Higgs decays in the $(y_\psi, \alpha)$ plane. Both decay modes are kinematically allowed.}
\label{fig:higgsconstraints}
\end{center}
\end{figure}

The impact of the non-standard Higgs decay constraints on the parameter space is exemplified in figure \ref{fig:higgsconstraints}, where we show
the allowed $(y_\psi, \alpha)$ combinations demanding for condition \eqref{eq:higgsinvcons} to be satisfied.
In this figure, we have varied $y_\psi$ in the interval $[10^{-3},10]$ and $\alpha$
within $[10^{-3},1]$, while identifying the SM-like Higgs boson with $h_2$ and setting  $m_{h_2} = 126$ GeV. We have moreover
kinematically allowed both $h_2 \rightarrow \psi\psi$ and $h_2 \rightarrow h_1 h_1$ decay modes, by choosing $m_\psi$ to vary within the range $[5,25]$ GeV
(i.e. the CDMS-Si compatible region) and $m_{h_1}$ within $[0.2, 63]$ GeV.

As a side remark let us note that interestingly, our findings show that the bound depicted in Fig.\ref{fig:higgsconstraints} is
very close to the one obtained if we only demanded $BR(h_2 \rightarrow h_1 h_1) < 0.3$. In other words, the bound is essentially set by the
decay mode of $h_2$ into two light scalars while the decay into two DM particles is less constraining. This feature might lead to the idea
that if $m_{h_1} > m_{h_2}/2$ or if we instead identified the light $h_1$ scalar with the SM Higgs boson, evading constraints from the LHC measurements could be far easier.
While this is generically true if we only consider the $h_2 \rightarrow h_1 h_1$ decay channel, in section \ref{sec:supressionmec} we will argue that resorting to such a choice would prohibit us to reproduce
the CDMS-Si result, avoid the constraints from XENON and LUX and satisfy LHC constraints at the same time.
In fact, such a choice would imply significantly increasing the
DM couplings to the Higgs boson in order to achieve the necessary scattering cross-sections, in contradiction with the limit   from the decay $h \rightarrow \psi\psi$ ~\cite{Djouadi:2011aa,Belanger:2013kya}.

A light scalar can also contribute to rare $\Upsilon$ and $B$ decays. In particular, new measurements of the process $\Upsilon\rightarrow \gamma \phi$ with the light scalar $\phi$ decaying into leptons and light mesons by the BELLE collaboration and precise measurements of the decay $B\rightarrow K\mu\mu$ by LHCb  can be used to constrain the light scalar couplings to Standard Model fermions for $\phi$ masses below 3 GeV. Recently, the authors of~\cite{Schmidt-Hoberg:2013hba} used the BELLE and LHCb data to extract the relevant limits for the couplings of a light scalar mediator to SM fermions in Higgs portal models of light dark matter.\footnote{Upon completion of our work, a similar study was also presented in \cite{Clarke:2013aya}.} 
In our notation, the authors find that the Higgs mixing angle $\alpha$ is bounded by $\sin\alpha \lesssim 7 \times 10^{-3}$ ( $\sim 9 \times 10^{-4}$) for $m_h=0.2 (2)$ GeV.
These constraints turn out to be extremely severe and indeed complementary to the invisible Higgs decay ones described before, since by comparing them with figure \ref{fig:higgsconstraints} we can deduce that for low $h_1$ masses they can cover a parameter space region that is otherwise fully allowed by the LHC results. In our analysis, we will impose the most stringent limit obtained in ~\cite{Schmidt-Hoberg:2013hba}, namely the LHCb result stemming from $B\rightarrow K\mu\mu$.
Additional constraints can come from contributions of $h_1$ to the anomalous magnetic moments of leptons. Using the formalism presented in
\cite{Goudelis:2011un}, we have checked that the constraints arising from these observables are satisfied throughout our treatment.

\subsection{Cosmological constraints and asymmetric dark matter}\label{sec:cosmo}

The Planck collaboration recently published its first results on the allowed dark matter abundance within the
$\Lambda$CDM cosmology~\cite{Ade:2013zuv}. In our analysis, we use the combined Planck+WMAP+BAO+High $L$ limit at
$3\sigma$,
\begin{equation}
\Omega_{CDM} h^2 = 0.1187 \pm  0.0051 .
\end{equation}
Note that in this work we consider asymmetric dark matter. This means that the relic density calculation introduces
an additional free parameter that can in principle be adjusted at will, namely the initial dark matter asymmetry.

\section{Suppression of Xenon detector constraints}\label{sec:supressionmec}
Having presented our model and the constraints it is subject to, we now turn to the mechanism that makes it possible to generate a visible signal in Si
detectors like CDMS while simultaneously evading bounds in Xe detectors.


\subsection{Analytical explanation}

All the DM direct detection experiments provide their results for the DM elastic scattering cross sections in terms of the ``normalized-to-nucleon'' cross-section, i.e. assuming isospin conserving couplings for neutrons and protons, $f_n=f_p$. However, in general DM can couple to neutrons and protons with different couplings, $f_n \neq f_p$. Moreover, if the signs of DM couplings to neutrons and protons are opposite, the corresponding contributions in a target nucleus can  cancel each other leading to a suppression of the interaction rate that depends on the number of neutrons and protons. When $f_n/f_p \simeq -0.7$, the scattering rate with the Xe target is most suppressed~\cite{Chang:2010yk, Feng:2011vu, Frandsen:2013cna}, thus weakening the constrains from XENON10~\cite{Angle:2011th},  XENON100~\cite{Aprile:2012nq} and LUX~\cite{Akerib:2013tjd}.


The effective Lagrangian for DM interaction with quarks contains both a vector and scalar interaction

\begin{equation}
{\cal L}=    f_q^V \bar{\psi} \gamma_\mu \psi    \bar{q}\gamma_\mu q   +  f_q^{h} \bar{\psi} \psi    \bar{q} q\,,
\end{equation}
where (see appendix ~\ref{app:interactions})
\begin{equation}
f_q^V=  \frac{g_\psi^Z(g_{qL}^Z+g_{qR}^Z)}{2 m^2_{Z}} + \frac{g_\psi^{Z_X} (g_{qL}^{Z_X}+g_{qR}^{Z_X})}{2 m^2_{Z_X}}
\end{equation}
and
\begin{equation}
f_q^h=  y_q y_\psi \frac{s_\alpha c_\alpha}{2} \left(  \frac{1}{m_{h_2}^2}-\frac{1}{m_{h_1}^2}  \right).
\end{equation}
The effective Lagrangian for nucleons has the same form as the one for quarks and  the effective couplings are related by means of
form factors. The scalar operator is interpreted as the
contribution of quark  $q$ to the nucleon mass $M_N$, and
$\langle N| m_q\overline{q} q|N\rangle=f^N_{Tq}M_N$, where the quark coefficients $f_{Tq}^N$ are computed from lattice calculations~\cite{Belanger:2013oya,Thomas:2012tg}. The vector interaction simply counts the number of valence quarks in the nucleon, thus,
\begin{equation}
f_p^V= 2 f_u^V+ f_d^V \;;  \;\;  f_n^V= f_u^V+ 2 f_d^V    \;;\;\; {\rm and} \;\;  f_{N}^h= \frac{M_N}{m_q}\sum_{q=u,d,c,s,t,b}  f_{Tq}^N f_q^h \;.
\end{equation}
The resulting amplitudes for DM (anti-DM) scattering on nucleons are given by  $f_N=f_N^h \pm f_N^V$  and the cross section for scattering off a point-like nucleus can be written as
\begin{equation}
\sigma_{\psi N}^0  =   \frac{4\mu^2}{\pi} \left [c ( Z f_p +(A-Z) f_n)^2 + \bar{c} ( Z \bar{f}_p +(A-Z) \bar{f}_n)^2\right]\,,
\label{eq:sig_anti}
\end{equation}
where $\mu$ is the DM-nucleus reduced mass $\mu = m_\psi m_{N}/(m_\psi + m_{N})$, $c=\frac{\rho_{\psi}}{\rho}$ ($\bar{c}=\frac{\rho_{\bar\psi}}{\rho}$) is the fractional contribution of the DM (anti-DM) component to the total local density, $\rho= \rho_{\psi}+\rho_{\bar\psi}$. We assume that $\rho_\psi/\rho=\Omega_\psi/\Omega$.
By inspecting \eqref{eq:sig_anti}, we can see that for symmetric dark matter ($\rho_\psi=\rho_{\bar\psi}$), the interference between the gauge and scalar contribution cancels out, since all crossed terms of the form $f_N^h f_N^V$ vanish.
On the other hand, the interference becomes maximal for asymmetric dark matter where one component completely dominates and the gauge and scalar contribution can be of the same order. In the asymmetric scenario, we can choose the couplings of the dominant component such that the scalar and gauge contributions are of the same order and interfere destructively. In that case, it is possible to approach the $f_n/f_p \simeq -0.7$ regime, where the couplings of DM to Xenon are suppressed.

Let us now consider the relative size of the different contributions.
When $r_X\equiv m_{Z_X}^2/m_Z^2 < 1$, as we consider here,  DM-nucleus scattering through gauge interactions should be dominated by mediation of the  $Z_X$ boson.
The vector interaction coupling between a $Z_X$ boson and a quark $q$ reads
\begin{align}
g_f^{Z_X} & = \frac{g_{fL}^{Z_X} + g_{fR}^{Z_X}}{2} \simeq \frac{e c_\xi t_\epsilon \sqrt{1-s_W^2} \left[ (8 s_W^2-4) Q + s_W^2 t_\epsilon^2 T_3 \right]}{8 s_W^2-4} + \mathcal{O}(r_X) \\ \nonumber
& \approx e c_\xi t_\epsilon c_W Q
\end{align}
since $t_\epsilon \ll 1$ in the small mixing limit. Thus in this limit, the effective coupling of DM to the neutron  via ${Z_X}$ interactions vanishes. The contribution due to Z exchange (suppressed by $r_X$) is on the other hand much larger for neutrons than protons since $ f_p^Z=(1-4 \sin^2\theta_W) f_n^Z$. The resulting vector amplitude nevertheless satisfies
 $f_p^{V} \gg f_n^{V}$ in the scenarios we will consider.
On the other hand, the effective couplings of DM to the proton and the neutron via scalar particles, $h_1$ and $h_2$, are almost the same: $f_p^{h_i} \simeq f_n^{h_i}$ since the interactions of $h_1$ and $h_2$ with a SM fermion $f$ are just proportional to the Yukawa coupling $y_f$ and $\sum f_{Tq}^p \approx \sum f_{Tq}^n$.
The neutron amplitude will therefore be dominated by the Higgs contribution with
$f_n \simeq f_n^{h_i} + f_n^{Z_X} \approx f_p^{h_i}$ while the proton amplitude is sensitive to both contributions. Consequently, one can find some region of  parameter space satisfying $f_n/f_p \approx f_p^{h_i}/(f_p^{h_i} + f_p^{Z_X}) \approx -0.7$. For this, one has to choose the parameters of the the gauge and scalar sector such that the gauge contribution is larger and of opposite sign than the scalar contribution, more precisely  $f_p^{h_i} \approx -0.4  f_p^{Z_X}$.  Here the sign of $f_{p, n}^{Z_X}$ is determined by the sign of the charge of DM under the U(1)$_X$ and we have chosen the sign  such that this condition is satisfied when DM dominates over anti-DM.

\subsection{Numerical demonstration}
In order to illustrate the previous arguments, we compute the normalized-to-nucleon scattering cross-section of DM off Si,Xe and Ge, which for a multi-isotope material reads
~\cite{Feng:2013vod}\begin{equation}
\sigma_{\psi N^Z}=\sigma_{\psi p} \left[    c \frac{\sum \eta_i \mu_{A_i}^2 (f_p Z+ f_n(A^i-Z))^2  }{\sum \eta_i \mu_{A_i}^2 f_p^2 }
+ \bar{c}  \frac{\sum \eta_i \mu_{A_i}^2 (\bar{f}_p Z+\bar{f}_n(A^i-Z))^2  }{\sum \eta_i \mu_{A_i}^2 \bar{f}_p^2 }
\right]\,,
\end{equation}
where $\eta_i$ and $\mu_{A_i}$ are the natural abundance and DM-nucleus reduced mass of the $i^{th}$ isotope and $c, \bar{c}$ are the relative abundances of $\psi$ and $\bar{\psi}$ respectively.
Note that in practice we have $\rho_\psi>>\rho_{\bar\psi}$ so that only the first term contributes.

\begin{figure}[t!]
\begin{center}
\includegraphics[width=0.49\textwidth]{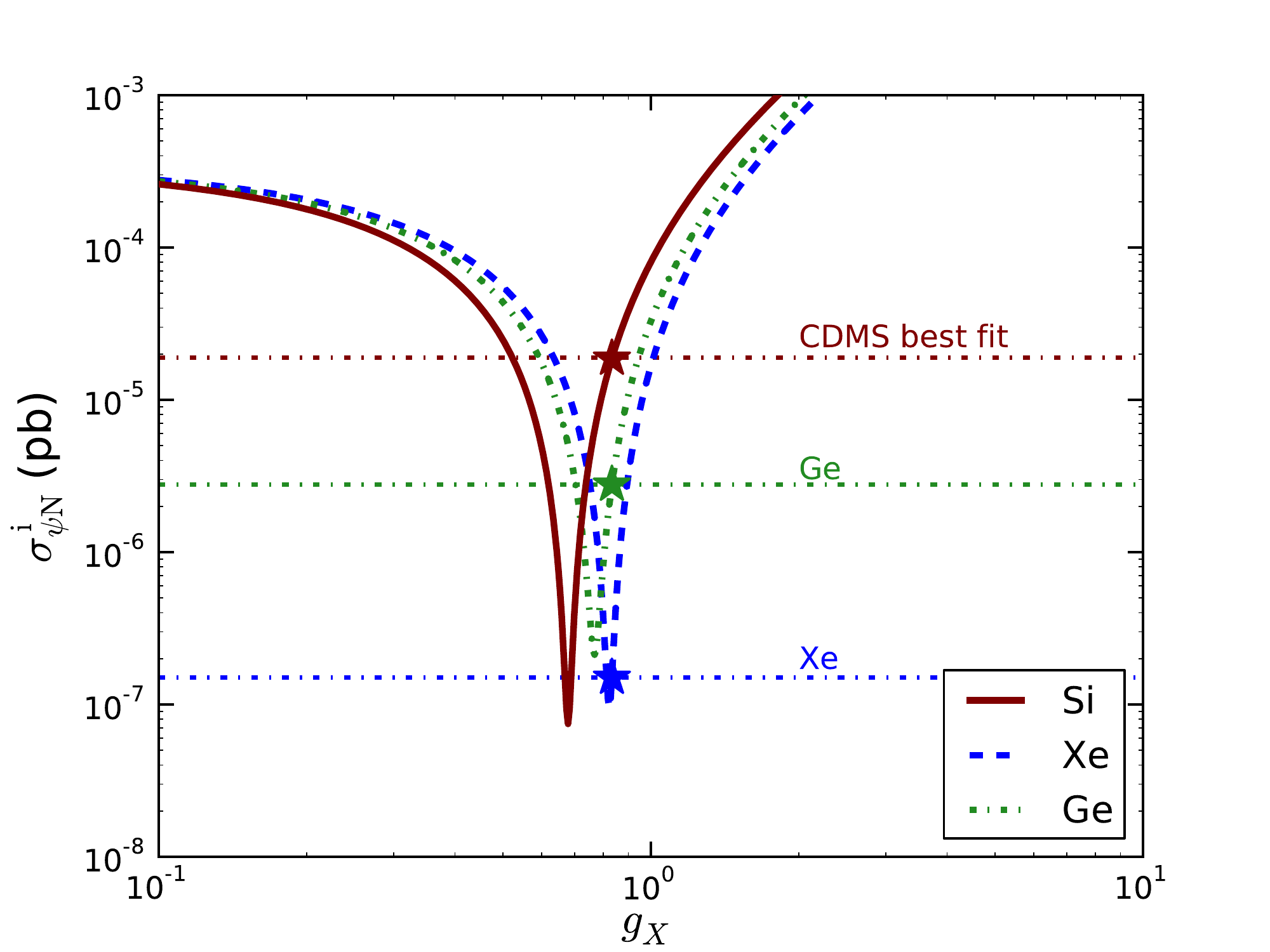}
\includegraphics[width=0.49\textwidth]{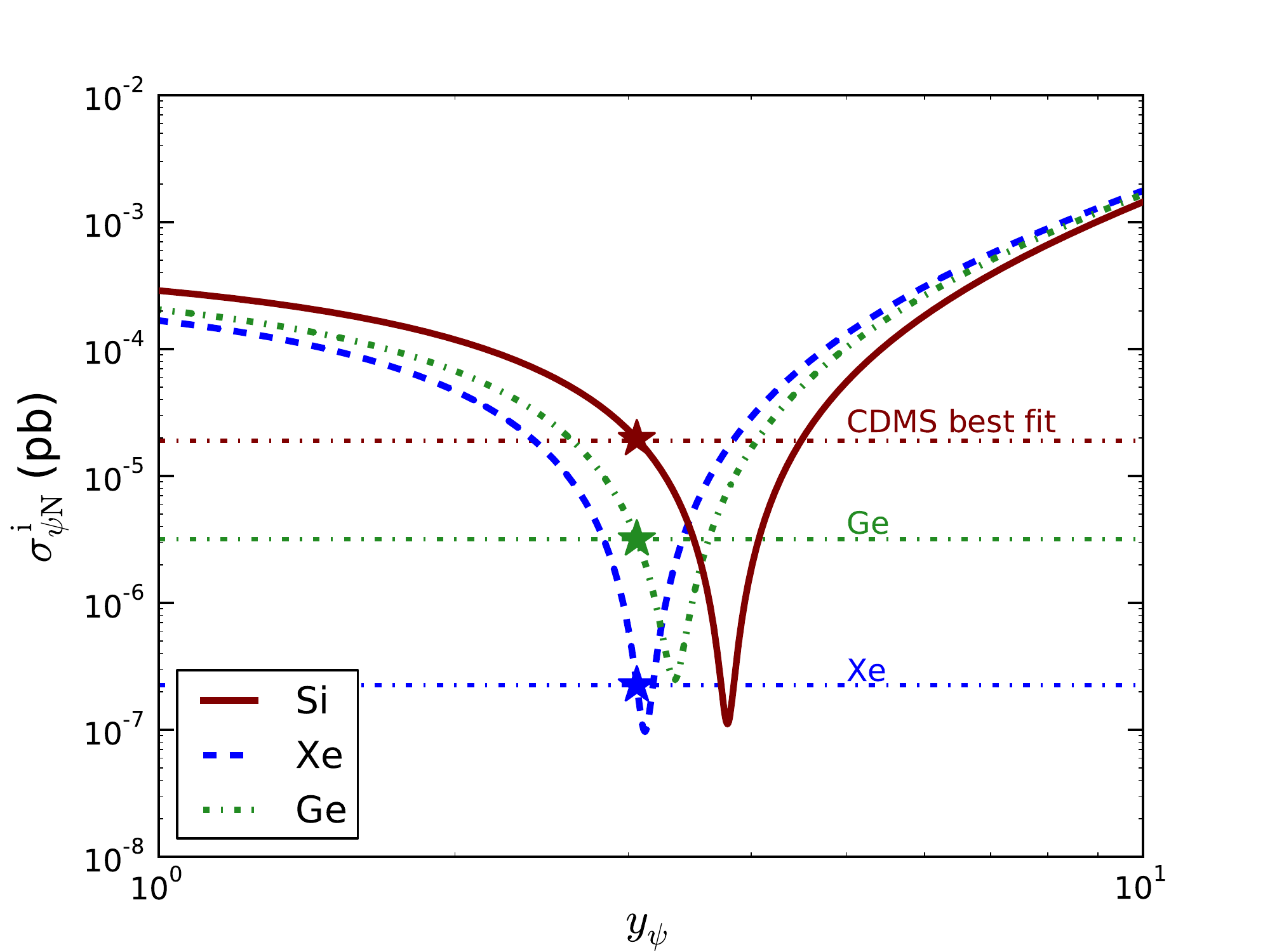}
\caption{The normalized-to-nucleon scattering cross-section off Si (brown, solid), Xe (blue, dashed) and Ge (green, dotted-dashed) as a function of $g_X$ (left)
and $y_\psi$ (right) for discrete choices of the other model parameters as described in the text. The horizontal lines show the CDMS-Si best-fit
cross-section and the corresponding scattering cross-section values for Ge and Xe. The stars correspond to points that reproduce the CDMS-Si excess
while having a strongly suppressed rate in Xenon, as shown in the figures.}
\label{fig:suppression1}
\end{center}
\end{figure}

The results are displayed  in Fig.\ref{fig:suppression1}.
Concretely, we fix all model parameters as shown in Table \ref{tab:suppression} and only vary the gauge
coupling $g_X$ (left panel) and the DM Yukawa coupling $y_\psi$ (right panel).
\renewcommand{\arraystretch}{1.1}
\begin{table}
\begin{center}
\begin{tabular}{|c|c|c|}
\hline \hline
Parameter  & Left panel & Right panel \\
 \hline
$m_Z$ & $91.1813$ & $91.1813$  \\
$m_{W}$ & $80.340$ & $80.340$    \\
$m_{Z_X}$ & $18$ & $18$  \\
$\rho$ & $0.9992$ & $0.9992$ \\
$m_\psi$ & $8.6$ & $8.6$  \\
$\epsilon$ & $7\times 10^{-3}$ & $7\times 10^{-3}$  \\
$m_{h_1}$ & $1$ & $1$  \\
$m_{h_2}$ & $126$ & $126$  \\
$\alpha$ & $8 \times 10^{-4}$ & $8 \times 10^{-4}$  \\
$g_X$ & - & $8.3 \times 10^{-1}$  \\
$y_\psi$ & $3.1$ & -  \\
\hline \hline
\end{tabular} \caption{Parameter values used in Fig.\ref{fig:suppression1}. All masses are in GeV.
\label{tab:suppression}}
\end{center}
\end{table}
\renewcommand{\arraystretch}{}
The dark matter mass is chosen to coincide with the best-fit
point as reported by the CDMS collaboration. The brown star in both panels shows the CDMS-Si best-fit cross-section, while the blue and green stars show the corresponding cross-section
values, for the same choice of  parameters, for Xe and Ge. The isotopic composition of all materials has been taken according to their natural abundances. Here, we have not imposed any
constraint on the depicted parameter combinations (although the CDMS-Si best-fit points satisfy all constraints discussed in section \ref{sec:psconstraints}),
since these figures are intended for illustration purposes.

From the figures, we can clearly see that the suppression mechanism can be extremely efficient, providing a maximal suppression factor for $\sigma_{\psi N}^{Xe}/\sigma_{\psi N}^{Si}$ up to ${\cal{O}}(100)$. The maximal suppression factor for the scattering cross-section off Ge  relative to Si is found to be of  ${\cal{O}}(10)$ for the
depicted points.  Note that a larger suppression factor can be obtained for other choices of parameters but the maximal suppression cannot be achieved at the same time for Ge and Xe.
The suppression factor for  Ar relative to Si is not quite as large as for Ge.
We should also point out that  the mechanism is quite sensitive to parameter variations, requiring very precise parameter combinations in order to be efficient. We therefore
do expect these results to be modified upon inclusion of radiative corrections, a study which goes well beyond the scope of the present work. Note however that electroweak corrections  have been shown to be large - albeit in  a different model -  only when the tree-level cross section is strongly suppressed~\cite{Klasen:2013btp}. We thus expect the general trend of our results to hold upon inclusion of radiative corrections.

\begin{figure}[t!]
\begin{center}
\includegraphics[width=0.49\textwidth]{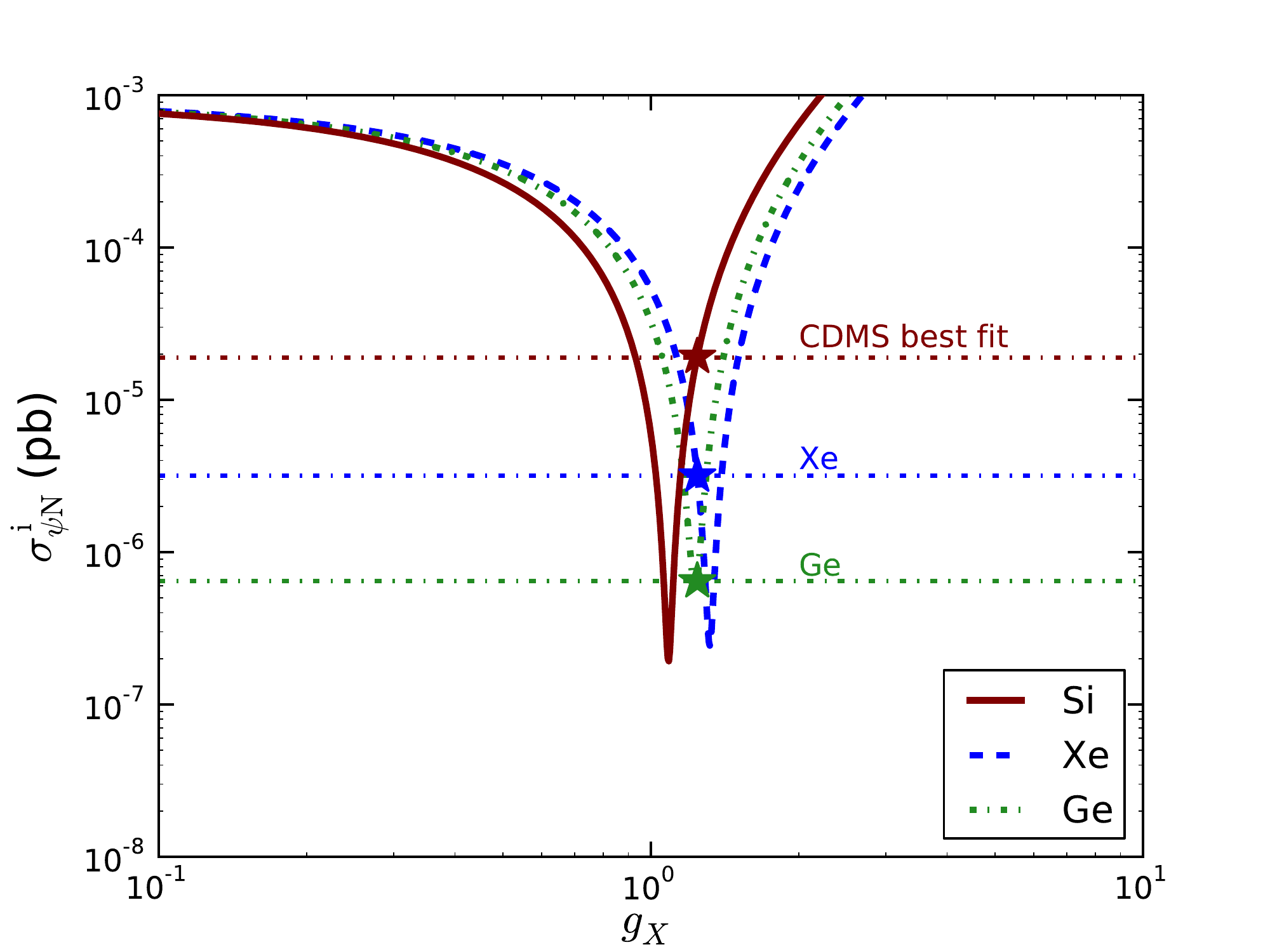}
\includegraphics[width=0.49\textwidth]{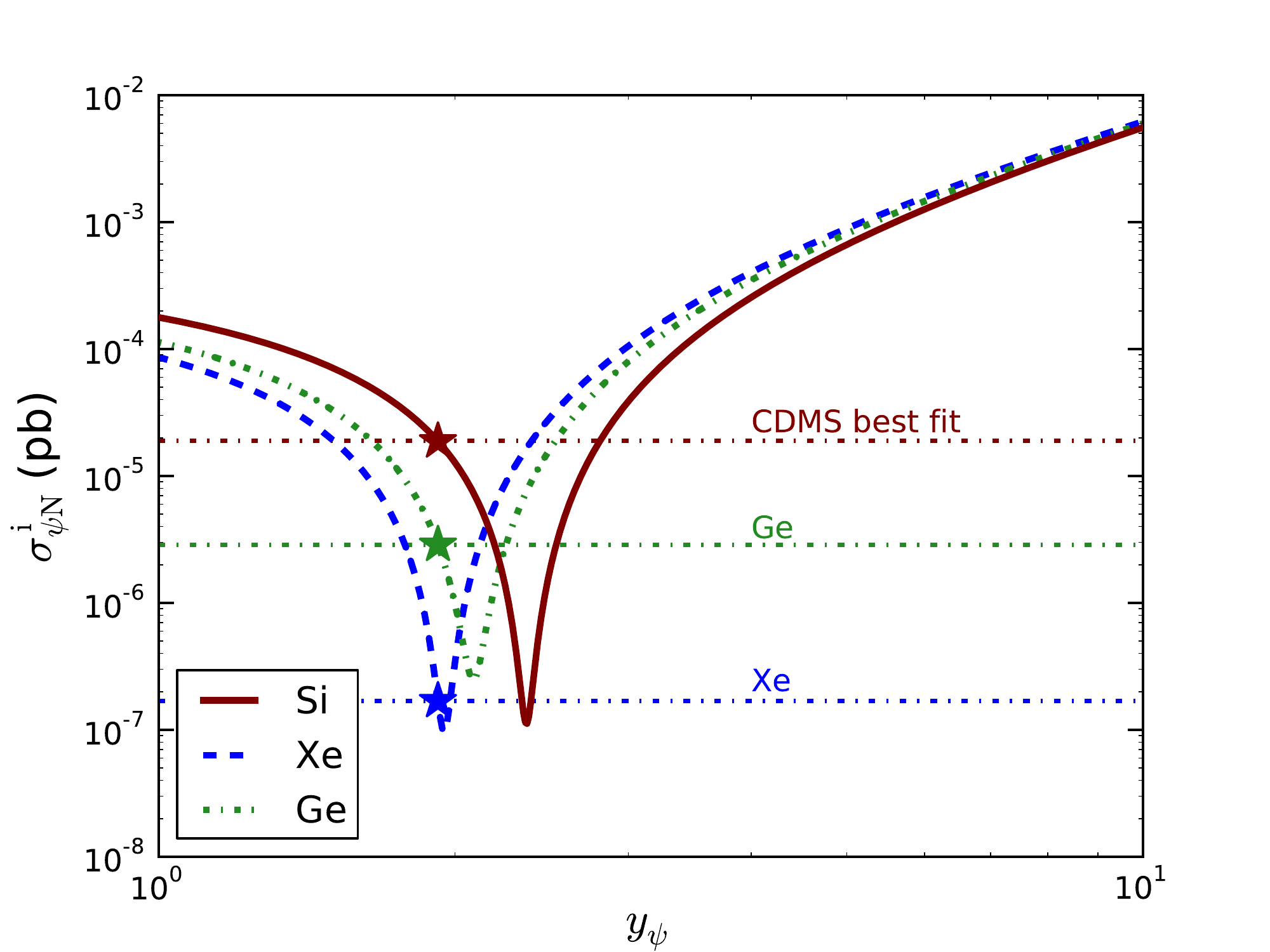}
\caption{Same labels as in Fig.~\ref{fig:suppression1}, but with different choices for the nucleon quark coefficients as described in the text}
\label{fig:suppression2}
\end{center}
\end{figure}

A further issue concerns the theoretical uncertainties tied to the values of the quark coefficients in the nucleon entering the scalar contribution and especially the $s$-quark coefficient, commonly denoted as $f_{Ts}$, which measures the strange quark content of the nucleon.
For the results displayed in Fig.~\ref{fig:suppression1}, we used the {\tt micrOMEGAs3} default values which correspond to $\sum_q f_{Tq}^p=0.28$~\cite{Belanger:2013oya}.  The impact of a larger value  $\sum_q f_{Tq}^p=0.47$ corresponding to the  default value of {\tt micrOMEGAs2.2}~\cite{Belanger:2008sj}, is shown in Fig.~\ref{fig:suppression2}. When the parameters of  the Higgs sector are fixed (left panel), the increase of the quark coefficient must be associated with an increase of the $Z_X$ contribution for a fixed value of $\sigma^{Si}_{\psi N}$, hence the larger value of $g_X$ at the CDMS best-fit point with respect to the one shown in Fig.~\ref{fig:suppression1}. This in turn implies a larger value for $f_n/f_p$ hence a less than optimal suppression factor for Xenon and an increased suppression factor for lighter nuclei such as Ge.
When the parameters of the gauge sector are fixed (right panel),
 the change in the quark coefficients can be compensated completely by a shift in the $h_1\bar\psi\psi$ coupling which determines the strength of the Higgs contribution. Hence the suppression factors for various nuclei are not affected.

A further important remark is that as we can clearly see, once the Xe cross-section is suppressed, the Si cross-section also undergoes a significant (although
milder) suppression. This means that the cross-section that we would get  if we were to switch off the
$Z_X$ ($g_X=0$) or Higgs ($y_\psi=0$) contributions
in the left  and right panel of figure~\ref{fig:suppression1} respectively would in fact be significantly larger than the CDMS-Si best fit. In other words,  large effective coupling
values are needed in order to be able to simultaneously reproduce the CDMS-Si cross-section while efficiently suppressing the
Xenon one.

This remark is of critical importance especially in the scalar sector of the model and is tightly connected to the discussion made at the end of
section \ref{sec:scalarsector}. The scalar-mediated scattering cross-section of a fermion off nucleons is governed
by essentially three factors :
the Yukawa-type couplings $h_i q \bar{q}$ and $h_i \psi\bar{\psi}$ (with the cross-section scaling quadratically with the
corresponding couplings) as well as the exchanged scalar mass (with the scattering cross-section scaling, for small $m_{h_i}$, roughly as $1/m_{h_i}^4$). In our
model, the $h_i q \bar{q}$ coupling is governed by the $\alpha$ angle and the usual quark Yukawa couplings, the $h_i \psi\bar{\psi}$ one is determined by $\alpha$ and $y_\psi$ whereas the masses $m_{h_i}$ are free parameters.
What we find in practice is that in order to achieve the necessary (large) scalar mediator contributions to the DM-nucleon scattering cross-section,
$m_{h_i}$ must be lighter than roughly  $5$ GeV or else the Higgs invisible branching ratio becomes prohibitively large.
Indeed, a heavier $h_i$ must be associated with a large value of  either $y_\psi$ or $\alpha$ thus leading to a large
$BR(h_i \rightarrow \psi\bar{\psi})$  and, if this mode is kinematically accessible,
$BR(h_2 \rightarrow h_1 h_1)$. We are therefore left with the choice of using the light Higgs mass
in order to achieve the necessary contributions to the scattering cross-section and
identifying $h_2$, the heavier scalar, with the SM-like Higgs boson of 126 GeV.
In what follows, we will therefore focus on the parameter space region where $h_1$ is very light.

However, as we already mentioned, this low-mass regime for $h_1$ is also severely constrained by bounds from flavour physics. Concretely, for $m_{h_1}$ in the
region $[0.2, 5]$ GeV,  $\sin\alpha$  cannot be larger than $7 \times 10^{-3}$. This small value is not detrimental to the DM-nucleon scattering cross-section, since it can be compensated by a large value  of the $y_\psi$ coupling, which however remains within perturbative limits. For example, as one can see
in Table \ref{tab:suppression}, for a light scalar mass of $1$ GeV and a scalar mixing angle $\alpha = 8\times 10^{-4}$, a coupling of $y_\psi \sim 3$ is
needed in order to achieve the required scattering cross-section values.
Note that the choice for the range of $m_{h_1}$ actually also illustrates an interesting example of the interplay of physics of two different scales.

\section{Results and discussion}\label{sec:scans}
In order to examine the parameter space of our setup, we have implemented the model in {\tt micrOMEGAs}~\cite{Belanger:2013oya} using the Feynrules package \cite{Christensen:2008py,Christensen:2009jx}. All observables have computed with {\tt micrOMEGAs} which relies on {\tt CalcHEP}~\cite{Pukhov:2004ca, Belyaev:2012qa} for the computation of cross-sections and decay widths. The relic density is computed assuming an initial  asymmetry in the DM abundance, $\Delta Y$, which is considered to be a free parameter.

Motivated by the  previous discussion, we performed extended scans over the parameter space of the model allowing the model parameters
to vary within the following intervals (all masses in GeV)
\begin{align}
91.1813 & <  m_Z < 91.1939 \nonumber\\
80.340 & < m_W < 80.430 \nonumber\\
0.9992 & < \rho < 1.0016\nonumber\\
0.003 & < \epsilon < 0.04  \nonumber\\
5 & < m_\psi < 25 \nonumber\\
2 m_\psi -7 & < m_{Z_X} < 2 m_\psi +7  \nonumber\\
0.005 & < y_\psi < 10\\
0.1 & < g_X < 10 \nonumber\\
123 & < m_{h_2} < 129 \nonumber\\
0.2 & < m_{h_1} < 5 \nonumber\\
1\times 10^{-4} & < \alpha < 5\times 10^{-3}\nonumber
\end{align}
whereas the dark matter asymmetry has been varied within the region $\Delta Y \in [1\times 10^{-11}, 1\times 10^{-10}]$. The parameter ranges have been chosen
so as to provide a full   parameter space coverage within the regions satisfying
the requirements presented in the previous sections. Note also that we have restricted the light scalar mass to be above $200$ MeV,
since going to lower masses would mean approaching the typical momentum transfer scale for DM-quark scattering, a regime in which the effective field
theory approach for DM-nucleon scattering breaks down.\footnote{Concretely, denoting the DM-quark scattering momentum transfer by $q$, the formulae implemented in {\tt micrOMEGAs} are formally valid in the limit $q \ll m_{h_1}$. One could indeed doubt the validity of this approximation for $m_{h_1} = 200$ MeV. We have verified that for this value of $m_{h_1}$ and a dark matter mass of $10$ GeV, the corrections induced to our estimates for the scattering cross-section off Xe are of ${\cal{O}}(5\%)$. The smallness of finite-$q$ effects is due to the fact that large momentum transfer events are suppressed both by the nuclear form factor and the Maxwell-Boltzmann velocity distribution.}

\begin{figure}[t!]
\begin{center}
\includegraphics[width=0.70\textwidth]{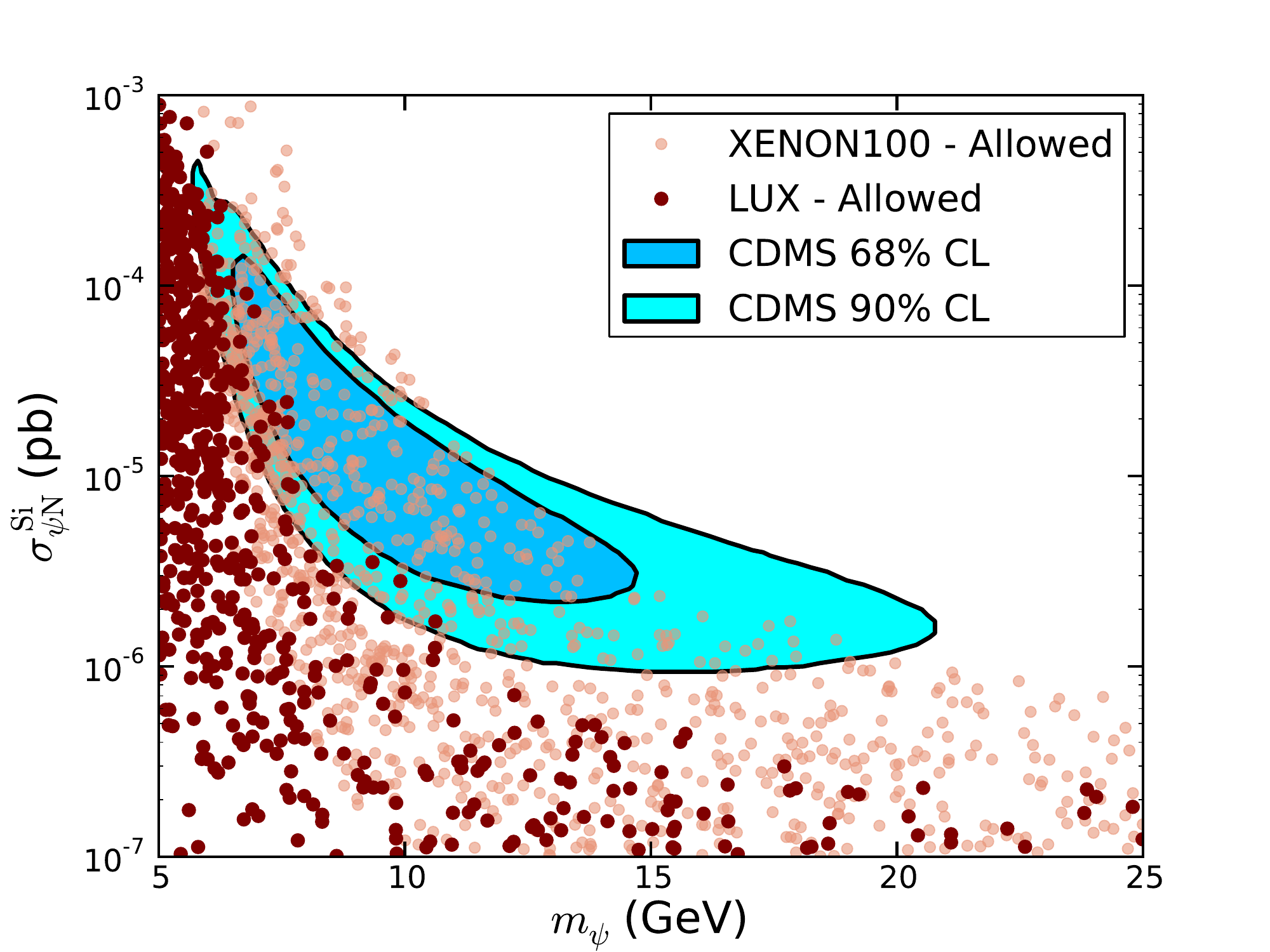}
\caption{Parameter space points in the $(m_\psi, \sigma_{\psi N}^{\rm Si})$ plane satisfying all experimental constraints and falling close
to the CDMS-Si compatible region for our choice of parameter ranges. Dark (pale) red points satisfy the LUX (XENON) bound.
The dark blue blob corresponds to the $68\%$ CL CDMS-Si compatible region whereas
the lighter one to the $90\%$ CL one.}
\label{fig:cdmsblob}
\end{center}
\end{figure}
\begin{figure}[t!]
\begin{center}
\includegraphics[width=0.70\textwidth]{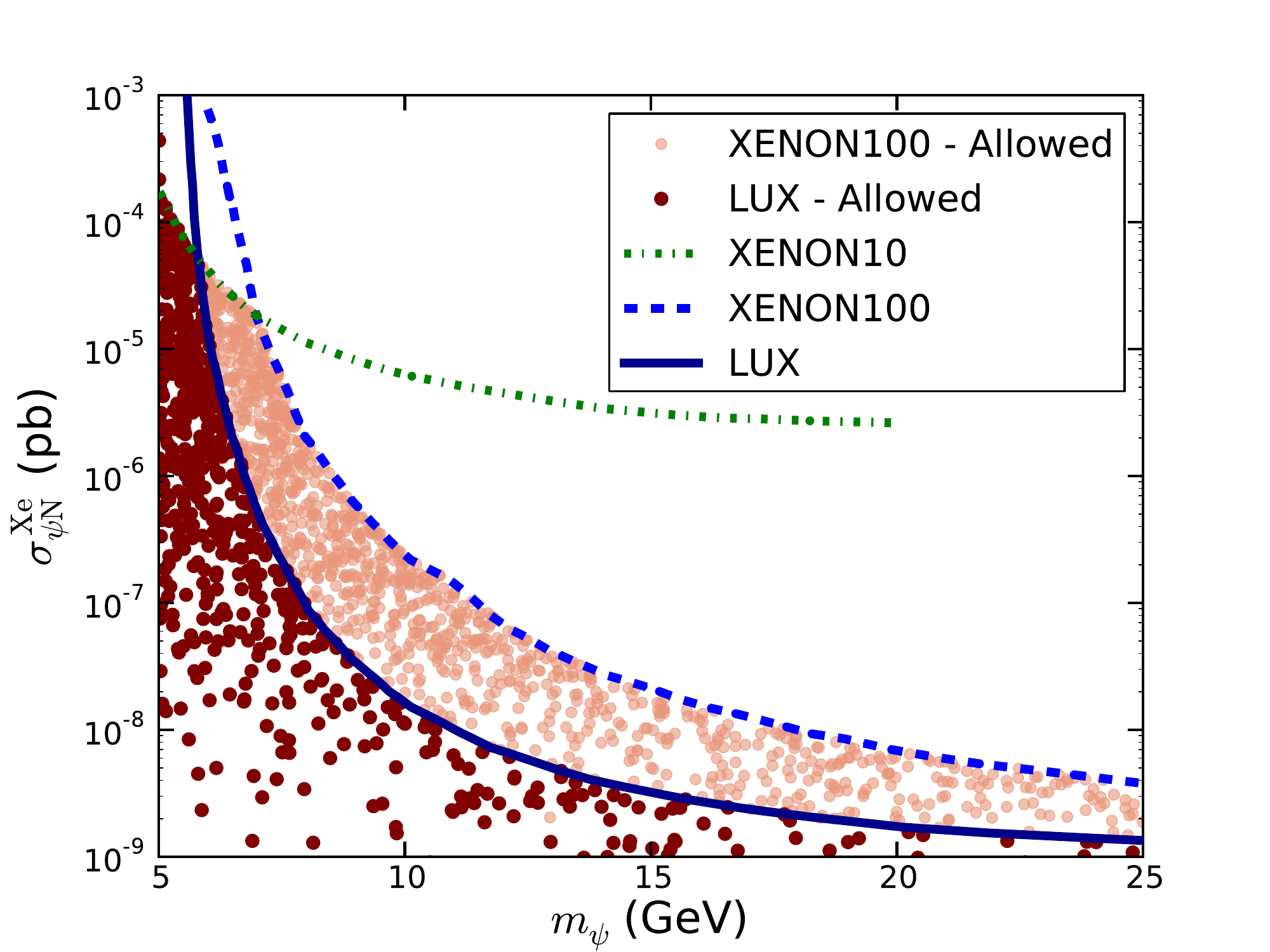}
\caption{Parameter space points in the $(m_\psi, \sigma_{\psi N}^{\rm Xe})$ plane satisfying all experimental constraints and falling close
to the CDMS-Si compatible region as in figure \ref{fig:cdmsblob}. The green dotted-dashed line corresponds to the XENON10 experimental bound, the light blue dashed one the the XENON100 one, while the darker blue solid line depicts the recent exclusion limits from the LUX experiment.}
\label{fig:XENON100comp}
\end{center}
\end{figure}

Our results for the DM scattering cross-section off Si are shown in figure \ref{fig:cdmsblob}, projected on the $(m_\psi, \sigma_{\psi N}^{Si})$ plane and displaying only the points for which $\sigma_{\psi N}^{Si}>1\times 10^{-7} {\rm pb}$. In the same figure, we also show the $68\%$ and $90\%$ CL regions that can fit the CDMS-Si excess.
All points depicted  respect the low-energy, collider, flavour physics and relic density constraints specified in section ~\ref{sec:psconstraints}
as well as the XENON10 and XENON100 bounds. The darker points also satisfy the $90\%$CL recent LUX bound
as explicited in figure~\ref{fig:XENON100comp}, where we project the same points  on the $(m_\psi, \sigma_{\psi N}^{\rm Xe})$ plane.

From these figures, we can see that with the simple setup we have adopted it is indeed possible to reconcile the recently observed CDMS-Si excess with the null searches from the XENON experiments, with the viable points of our parameter space covering essentially the full CDMS-compatible region.
However the LUX exclusion bound leaves only a narrow strip in the  CDMS-Si compatible region  corresponding to
$m_\psi < 10$ GeV.

\begin{figure}[t!]
\begin{center}
\includegraphics[width=0.70\textwidth]{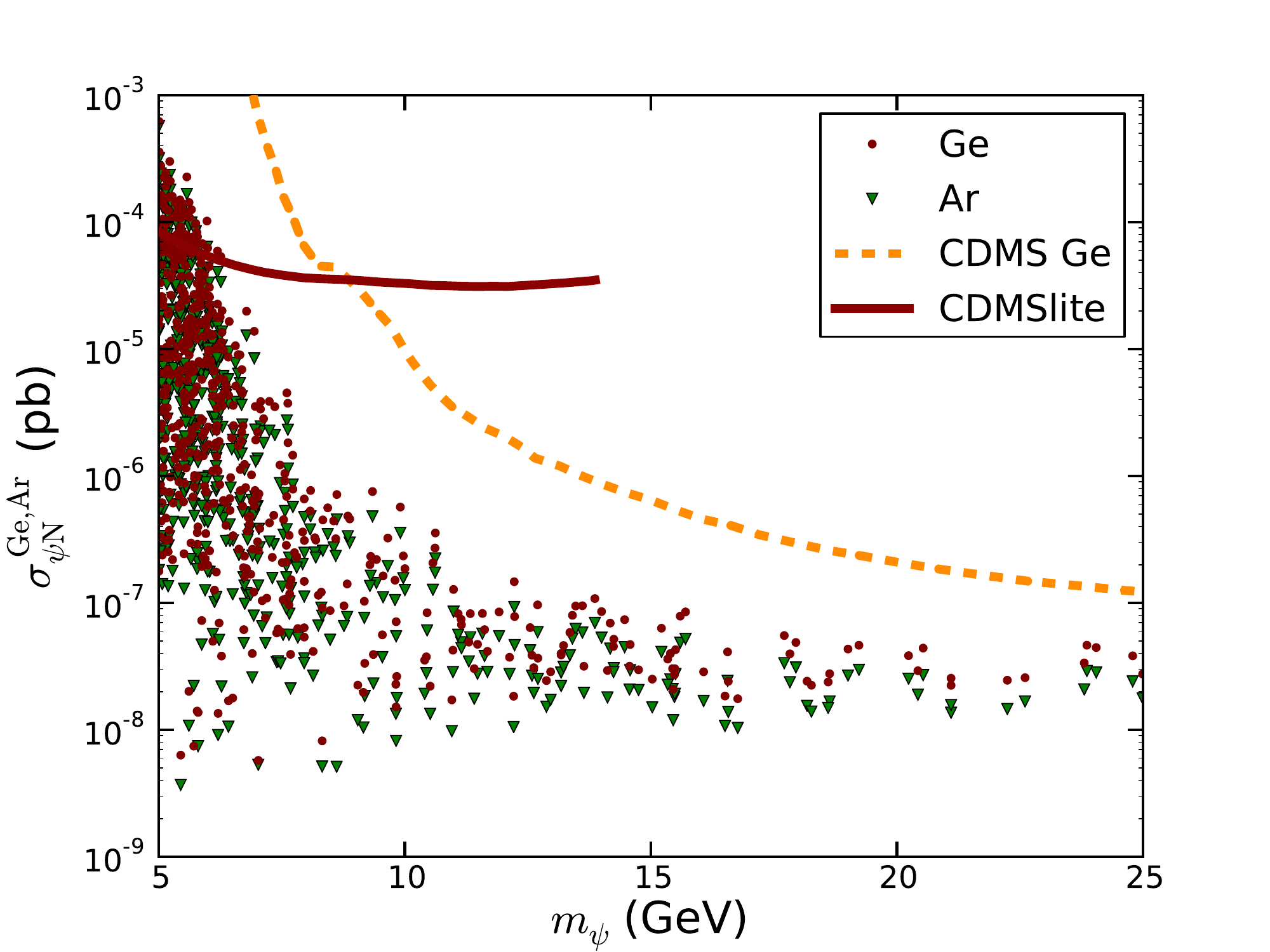}
\caption{Parameter space points in the $(m_\psi, \sigma_{\psi N}^{\rm Ge, Ar})$ plane satisfying all experimental constraints and falling close
to the CDMS-Si compatible region as in figure \ref{fig:cdmsblob}. $(m_\psi, \sigma_{\psi N}^{\rm Ge})$ values are depicted in brown circles
while $(m_\psi, \sigma_{\psi N}^{\rm Ar})$ in green triangles. The dark red solid line depicts the exclusion bounds coming from the CDMS-lite analysis
whereas the orange dashed one to the CDMS-Ge one. Both bounds should only be compared to the $(m_\psi, \sigma_{\psi N}^{\rm Ge})$ points.}
\label{fig:GeAllOK}
\end{center}
\end{figure}

For completeness, in Fig.\ref{fig:GeAllOK} we also show the same results for the scattering cross-section off Germanium (brown circles) and Argon (green triangles).
Typically, these cross sections
are suppressed by a factor $10$ for Ge as compared with Si, thus most points satisfy the CDMS-Ge exclusion, with only a few points at low mass exceeding the limit obtained recently in the CDMS-lite study~\cite{Agnese:2013lua}.
Moreover, we find some points in the region favoured by CoGeNT corresponding to $\sigma^{Ge}_{\psi N}\sim 2-4 \times 10^{-5}$~pb.
In general, the suppression factor for Ar is a factor of two weaker than  for Ge, especially when a near maximal suppression factor is required for Xe. For instance, this is the case for  points with $m_\psi>10$ GeV. However, the suppression factor can be larger for Ar than for Ge. This occurs, for example, for very light DM ($m_\psi <7$~GeV) where $f_n/f_p$ can differ significantly from -0.7 since in this mass range the limit from Xenon detectors is relaxed. In particular,  a value close to  $f_p/f_n=-0.82$ which leads to the maximal  suppression for Ar can satisfy all the constraints.

Our results clearly demonstrate the complementarity of dark matter detectors operating with different materials, since the large suppression of the scattering cross-section that might occur in Xe relative to Si will necessarily be milder for lighter nuclei such as Ar and Ge.
This in turn shows the relevance of an increased sensitivity in detectors with light nuclei for a thorough test of models with isospin-violating interactions,
although in the foreseeable future the region of parameter space compatible with CDMS-Si will best be probed by increasing the sensitivity of Xenon detectors. The recent improvement of the relevant exclusion limit with a Xenon detector, LUX, has indeed closed a large portion of the CDMS-Si allowed parameter space.

\begin{figure}[t!]
\begin{center}
\includegraphics[width=0.70\textwidth]{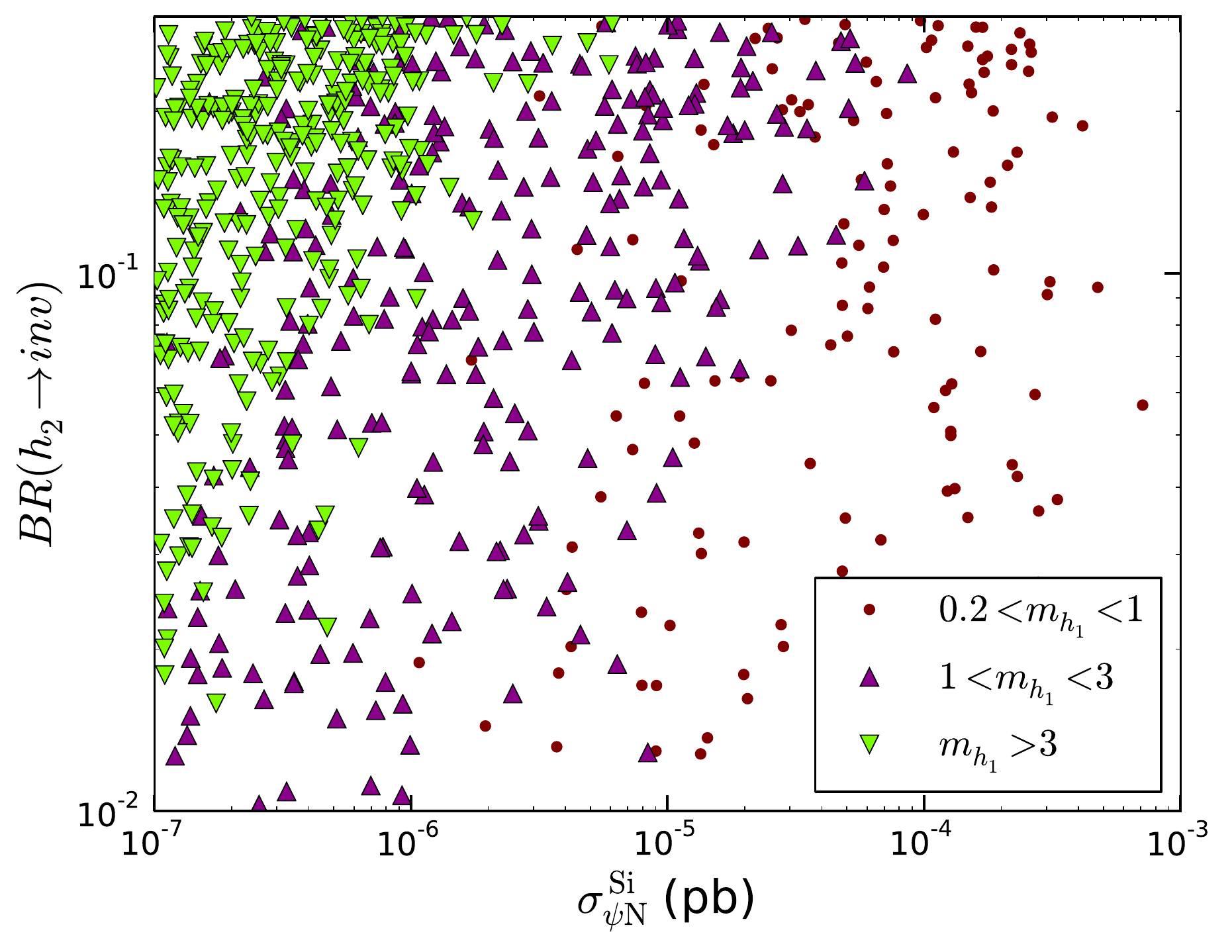}
\caption{The SM-like Higgs branching ratio into invisible final states against the normalized-to-nucleon scattering cross-section off Si for
parameter space points satisfying all experimental constraints and falling close to the CDMS-Si compatible region for our choice of parameter ranges.
Brown circles correspond to points for which the light Higgs mass is between $0.2$ and $1$ GeV, violet upwards triangles to points where
$1 < m_{h_1} < 3$ GeV and green downwards triangles to $m_{h_1} > 3$ GeV.}
\label{fig:SiBRinv}
\end{center}
\end{figure}

Interestingly, dark matter searches are not the only source of information for our model. In Figure~\ref{fig:SiBRinv}, we show the predictions of this model for the SM-like Higgs ($h_2$) invisible decay branching ratio as a function of the scattering cross-section off Si, for the points depicted in Fig.\ref{fig:cdmsblob} that satisfy all experimental constraints except the recent LUX bound. In order to illustrate moreover the correlation between the invisible Higgs branching ratio and the light ($h_1$) Higgs mass, we delineate three regions for the latter: $0.2 < m_{h_1} < 1$ GeV (brown circles), $1 < m_{h_1} < 3$ GeV (violet upwards triangles) and $m_{h_1} > 3$ GeV (green downwards triangles). This figure is strongly related to the discussion on the possible values of the light Higgs mass in order to reproduce CDMS-Si while evading all other constraints. We see that for relatively large values of $m_{h_1}$ the required cross-section can be barely reached, whereas in the cases where this is possible the corresponding SM Higgs invisible branching ratios are large enough so that they should be accessible at the next LHC run once improved measurements of the Higgs decay properties are performed. The lower $h_1$ mass regime is however more elusive in Higgs studies. We expect that improved analyses on $B$ meson decays coming from LHCb should provide interesting information for this mass range. Concretely, if the Higgs mixing angle $\alpha$ is further pushed towards lower values, then light Higgs masses above roughly $1$ GeV should become inefficient in providing such large DM-nucleon scattering cross sections since the required Yukawa coupling values would start entering the deep non-perturbative regime.

\section{Conclusion}\label{sec:conclusions}

In this work, we have shown that a minimal U(1) extension of the standard model with a Dirac fermion dark matter and a light singlet could be compatible with
the excess of events observed in CDMS-Si, while avoiding the strong constraints from the LUX experiment, by yielding isospin-violating interactions between DM and nucleons.
In this model, the relic DM density is linked to a DM/anti-DM asymmetry in the early Universe which, being of the same order as the baryon/anti-baryon asymmetry, could have a similar origin. The present day DM asymmetry is crucial for generating the isospin violating interactions as it provides an interference between the scalar and vector boson contribution in DM elastic scattering on nucleons. The scalar sector of the model can be tested further at colliders both with precise measurements of the Higgs properties - in particular the invisible width - and improved measurements of rare B-decays. The new light gauge boson and Dirac fermion are more elusive at colliders as they couple to SM particles only through small mixing effects.

When presenting our results, we have concentrated on the region of parameter space that contains a light DM Dirac fermion with a large direct detection rate in Si and a strongly suppressed one for Xe. However, we stress that the mechanism we have proposed for producing isospin-violating interactions can also be associated with  lower cross sections on Si, with heavier dark matter candidates and with different suppression factors on various nuclei depending on the region of parameter space under consideration. Therefore, irrespective of the fate of the present hints of DM in direct detection and of the details of this specific model, this work stresses the importance of searching for dark matter with detectors made of different (both light and heavy) nuclei. In the future, confronting signals obtained with different detectors could thus provide extremely useful information on the properties of the dark matter candidate.

\section{Acknowledgments}
We thank Pasquale Serpico and Aoife Bharucha for useful discussions.
This work was supported in part by the French ANR, project DMAstroLHC and by  the LIA-TCAP of CNRS.
A.P. was supported by the Russian foundation for Basic Research, grant RFBR-12-02-93108-CNRSL-a.
J.-C.P. is supported by Basic Science Research Program through the National Research Foundation of Korea (NRF) funded by the Ministry of Education (2011-0029758) and (2013R1A1A2061561).

\appendix

\section{Interactions in the physical basis}\label{app:interactions}

Let us list all the interaction vertices of the physical $W, Z$ and $Z_X$ gauge bosons relevant for our analysis. In order to describe
the interaction vertices of $W, Z$ and $Z_X$, let us define the various couplings, $g$'s, as follows:
\begin{align}\label{g's}
 {\cal L} & = W_\mu^+\, g_f^W [ \bar{\nu} \gamma^\mu P_L e
                        + \bar{u} \gamma^\mu P_L d ] + c.c. \nonumber \\
  & + Z_\mu \left[ g^Z_{fL}\, \bar{f} \gamma^\mu P_L f
                          + g^Z_{fR}\, \bar{f} \gamma^\mu P_R f
       + g^Z_{\psi}\, \bar{\psi}\gamma^\mu\psi \right] + g_W^Z [[Z W^+ W^-]] \nonumber \\
  & + Z_{X \mu} \left[ g^{Z_X}_{fL}\, \bar{f} \gamma^\mu P_L f
                + g^{Z_X}_{fR}\, \bar{f} \gamma^\mu P_R f
       + g^{Z_X}_{\psi}\, \bar{\psi_1}\gamma^\mu\psi \right] + g_W^{Z_X} [[Z_X W^+ W^-]] \nonumber \\
  & + h_1 \left[  g^{h_1}_{ZZ}\, Z_\mu Z^\mu + g^{h_1}_{XX} Z_{X \mu} Z_X^\mu + g^{h_1}_{XZ} Z_{X \mu} Z^\mu \right]
  \nonumber \\
  & + h_2 \left[  g^{h_2}_{ZZ}\, Z_\mu Z^\mu + g^{h_2}_{XX} Z_{X \mu} Z_X^\mu + g^{h_2}_{XZ} Z_{X \mu} Z^\mu \right] \,.
\end{align}
These redefined couplings expressed by the physical observables (unhatted parameters) can be obtained from the appendix of Ref.~\cite{Chun:2010ve}:
\begin{align}
g_f^W & = - {e\over \sqrt{2} s_W}
 \left(1-{\omega \over 2(1-t_W^2)}\right)\,, \nonumber \\
g^Z_{fL} & = -{e\over c_{{W}} s_{{W}} }\, c_\xi \,
         \left\{ T_3 \left[1+ {\omega \over 2}\right]
         - Q \left[s_{{W}}^2 + \omega \left( {2 - t_W^2
         \over 2( 1- t_W^2) }\right) \right] \right\}\,, \nonumber\\
g^Z_{fR} & = {e\over c_{{W}} s_{{W}} }\, c_\xi \,
         Q \left[s_{{W}}^2 + \omega \left( {2 - t_W^2
         \over 2( 1- t_W^2) }\right) \right]\,,    \nonumber\\
g^Z_\psi & = g_X {s_\xi\over c_\epsilon}\,, \nonumber
\end{align}
\begin{align}
g^{Z_X}_{fL} & = - {e\over c_{{W}} s_{{W}} }\, {c_\xi} \left\{
         T_3 \left[ s_W t_\epsilon - t_\xi + {1\over 2}\,
         \omega \left(t_\xi + { s_W t_W^2 t_\epsilon\over 1 - t_W^2} \right) \right]
         \right. \nonumber\\
  & ~~~~~~~~~~~~~~\left. + Q \left[ s_W^2 t_\xi - s_W t_\epsilon
         + {1\over 2}\, t_W^2 \omega \left({t_\xi -s_W t_\epsilon \over 1-t_W^2 } \right) \right]
         \right\}\,, \nonumber\\
g^{Z_X}_{fR} & = -{e\over c_{{W}} s_{{W}} }\, {c_\xi}\, Q \left[ s_W^2 t_\xi - s_W t_\epsilon
         + {1\over 2}\, t_W^2 \omega \left({t_\xi -s_W t_\epsilon \over 1-t_W^2 } \right)
         \right]\,,  \nonumber\\
g^{Z_X}_\psi & = g_X {c_\xi\over c_\epsilon}\,, \nonumber
\end{align}
\begin{align}
g_W^Z &= {e \over t_{{W}}}\, c_\xi \left(1-{\omega \over 2( c_W^2-s_W^2)} \right)\,, \nonumber\\
g_W^{Z_X} &= - {e \over t_{{W}}}\, s_\xi \left(1-{\omega \over 2( c_W^2-s_W^2)} \right)\,, \nonumber
\end{align}
\begin{align}
g^{h_1}_{ZZ} &= -s_\alpha {m_{{Z}}^2\over v}\, c_\xi^2\, (1 + \omega)\,, \nonumber\\
g^{h_1}_{XX} &= -s_\alpha {m_{{Z}}^2\over v}\, c_\xi^2\, \left[ t_\xi^2 + s_W^2 t_\epsilon^2
               - \omega \left(2 + t_\xi^2 - { s_W^2 t_W^2 t_\epsilon^2 \over
               1 - t_W^2} \right) \right]\,, \nonumber\\
g^{h_1}_{XZ} &= -s_\alpha {m_{{Z}}^2\over v}\, c_\xi^2\, 2 \left[ 2s_W t_\epsilon
               - t_\xi + \omega \left( t_\xi + {s_W t_W^2 t_\epsilon
               \over 1- t_W^2} \right) \right]\,, \nonumber
\end{align}
\begin{align}
g^{h_2}_{ZZ} &= c_\alpha {m_{{Z}}^2\over v}\, c_\xi^2\, (1 + \omega)\,, \nonumber\\
g^{h_2}_{XX} &= c_\alpha {m_{{Z}}^2\over v}\, c_\xi^2\, \left[ t_\xi^2 + s_W^2 t_\epsilon^2
               - \omega \left(2 + t_\xi^2 - { s_W^2 t_W^2 t_\epsilon^2 \over
               1 - t_W^2} \right) \right]\,, \nonumber\\
g^{h_2}_{XZ} &= c_\alpha {m_{{Z}}^2\over v}\, c_\xi^2\, 2 \left[ 2s_W t_\epsilon
               - t_\xi + \omega \left( t_\xi + {s_W t_W^2 t_\epsilon
               \over 1- t_W^2} \right) \right]\,,
\label{eq:interaction_couplings}
\end{align}
where $\omega =  s_W t_\xi t_\epsilon \simeq -(1 - t_W^2 ) (\rho-1) \sim \mathcal{O} (10^{-3}) $.
In addition, we obtain the couplings among three Higgs bosons,
$h_2 h_1 h_1$ and $h_1 h_2 h_2$, that read
\begin{align}
& g_{h_2 h_1 h_1} = \frac{3}{2} s_\alpha c_\alpha (\lambda_S v_S c_\alpha + \lambda_H v s_\alpha)
+ \lambda_{SH} [v_S s_\alpha (s^2_\alpha - 2 c^2_\alpha)
+ v c_\alpha (c^2_\alpha - 2 s^2_\alpha)]\,, \label{h211Couplings} \\
& g_{h_1 h_2 h_2} = \frac{3}{2} s_\alpha c_\alpha (\lambda_S v_S s_\alpha - \lambda_H v c_\alpha)
+ \lambda_{SH} [v_S c_\alpha (c^2_\alpha - 2 s^2_\alpha)
- v s_\alpha (s^2_\alpha - 2 c^2_\alpha)]\,.  \label{h122Couplings}
\end{align}
%

\bibliographystyle{JHEP}

\providecommand{\href}[2]{#2}\begingroup\raggedright\endgroup

\end{document}